\documentclass[runningheads]{moh}
\usepackage[utf8]{inputenc}
\usepackage[T1]{fontenc}
\usepackage[a4paper, total={6.5in, 9.5in}]{geometry}
\usepackage{multicol}
\usepackage{tabularx}
\usepackage{graphicx}
\usepackage{float}
\usepackage{caption}
\usepackage{subcaption}
\usepackage{multirow}
\PassOptionsToPackage{hyphens}{url}\usepackage{hyperref}
\usepackage{amsmath,amssymb,amsfonts}
\usepackage{algorithmic}
\usepackage{textcomp}
\usepackage{array}
\newcolumntype{L}{>{\centering\arraybackslash}m{4.9cm}}
\usepackage{verbatimbox}
\usepackage{varwidth}

\providecommand{\keywords}[1]{\textbf{\textit{Index terms---}} #1}

\title{Towards a Standard Feature Set for \\Network Intrusion Detection System Datasets}
\titlerunning{Towards a Standard Feature Set of NIDS Datasets}
\author{Mohanad Sarhan\inst{1}\and
Siamak Layeghy\inst{1}\and
Marius Portmann\inst{1}}
\authorrunning{Sarhan et al.}
\institute{University of Queensland, Brisbane QLD 4072, Australia 
\\
\email{m.sarhan@uq.net.au};
\email{siamak.layeghy@uq.net.au};
\email{marius@ieee.org}\\
}
\date{\vspace{-5ex}}

\begin{document}

\maketitle

\begin{abstract}
Network Intrusion Detection Systems (NIDSs) are important tools for the protection of computer networks against increasingly frequent and sophisticated cyber attacks. Recently, a lot of research effort has been dedicated to the development of Machine Learning (ML) based NIDSs. As in any ML-based application, the availability of high-quality datasets is critical for the training and evaluation of ML-based NIDS. One of the key problems with the currently available NIDS datasets is the lack of a standard feature set. The use of a unique and proprietary set of features for each of the publicly available datasets makes it virtually impossible to compare the performance of ML-based traffic classifiers on different datasets, and hence to evaluate the ability of these systems to generalise across different network scenarios. To address that limitation, this paper proposes and evaluates standard NIDS feature sets based on the NetFlow network meta-data collection protocol and system. 
We evaluate and compare two NetFlow-based feature set variants, a version with 12 features, and another one with 43 features. For our evaluation, we converted four widely used NIDS datasets (UNSW-NB15, BoT-IoT, ToN-IoT,  CSE-CIC-IDS2018) into new variants with our proposed NetFlow based feature sets.  
Based on an Extra Tree classifier, we compared the classification performance of the NetFlow-based feature sets with the proprietary feature sets provided with the original datasets. 
While the smaller feature set cannot match the classification performance of the proprietary feature sets, the larger set with 43 NetFlow features, surprisingly achieves a consistently higher classification performance compared to the original feature set, which was tailored to each of the considered NIDS datasets.  
%
%
The proposed NetFlow-based standard NIDS feature set, together with four benchmark datasets, made available to the research community, allow a fair comparison of ML-based network traffic classifiers across different NIDS datasets. We believe that having a standard feature set is critical for allowing a more rigorous and thorough evaluation of ML-based NIDSs and that it can help bridge the gap between academic research and the practical deployment of such systems.


\end{abstract}


\section{Introduction}
Network Intrusion Detection Systems (NIDSs) aim to detect network attacks and to preserve the three principles of information security: confidentiality, integrity, and availability \cite{modi2012bayesian}. Signature-based NIDSs match attack signatures to observed traffic, giving a high detection accuracy to known attacks. However, these systems are unable to detect previously unseen (zero-day) attacks or new variants of known attacks. Therefore, researchers have investigated anomaly-based NIDSs that focus on matching attack behaviours and patterns \cite{garcia2009anomaly}. Machine Learning (ML), a sub-field of artificial intelligence, is capable of learning and extracting complex network attack patterns that may threaten computer networks if undetected  \cite{6779523}. All network intrusions generate a unique set of security events, that would aid in their classification process. These identifying patterns can be extracted from network traffic in the form of data features. To generate a dataset, corresponding data features form network data flows that are ideally labelled with an attack or a benign class to allow for a supervised ML methodology. 

Real-world network flow datasets with labels that identify attack and benign flows are challenging to obtain, mainly due to security and privacy concerns.  Therefore, researchers have designed network test-beds to generate synthetic datasets that consist of labelled network data flows \cite{SHIRAVI2012357}. The data flows are made of several network features that are often preselected based on the authors' domain knowledge and available extraction tools. As a result, the currently available NIDS datasets are very distinct in terms of their feature sets and therefore the security events represented by the data flows. Due to the great impact of data features on the performance of ML models \cite{binbusayyis_vaiyapuri_2019}, the evaluation of the proposed ML-based NIDSs are often unreliable when tested on multiple datasets using their original feature sets. Finally, as certain network data features require a complex and deep packet inspection, the computational complexity of feature extraction and processing is not feasible


The importance of having a standard feature set for all datasets is paramount. It will facilitate a fair and reliable evaluation of proposed ML models across various network environments and attack scenarios. This also enables an evaluation of the generalisability of the model, and hence its performance when deployed in practical network scenarios. Moreover, a standard feature set will ensure that the security events and network information presented by NIDS datasets are the same and in a controlled manner. NetFlow is an industry-standard protocol for network traffic collection \cite{claise2004cisco}.Its practical and scalable deployment properties are capable of enhancing the deployment feasibility of ML-based NIDSs.  NetFlow features are capable of presenting key security events that are crucial in the identification of network attacks. Therefore, we believe that applying NetFlow-based features in the design of a universal feature set will facilitate the successful deployment of ML-based NIDS in practical network scenarios.

Four widely used NIDS datasets, referred to as UNSW-NB15 \cite{moustafa-slay-2015}, BoT-IoT \cite{DBLP:journals/corr/abs-1811-00701}, ToN-IoT \cite{fesz-dm97-19},  and CIC-CSE-IDS2018 \cite{sharafaldin-habibi-lashkari-ghorbani-2018} have been converted into a common basic NetFlow-based feature set \cite{sarhan2020netflow}. The NetFlow datasets address some of the current research issues by applying a common feature set across multiple datasets. However, due to the insufficient security information represented by the basic NetFlow feature set, the ML models lead to limited detection accuracy, in particular when performing multi-class experiments. Therefore, this paper proposes an extended NetFlow feature set as the standard version to be used in future NIDS datasets. As part of its evaluation, the features have been extracted and labelled from four well-known datasets. The datasets generated are named NF-UNSW-NB15-v2, NF-BoT-IoT-v2, NF-ToN-IoT-v2, NF-CSE-CIC-IDS2018-v2 and NF-UQ-NIDS-v2, and have been made publicly available for research purposes \cite{netflow_datasets_2020}.

This paper explores two variants of NetFlow-based feature sets along with their proprietary feature sets. The rest of the paper is organised as follows. Existing NIDS datasets and their limitations are discussed in Section \ref{lm}. Section \ref{NF} motivates the case for having a standard and a common feature set in NIDS datasets. It also illustrates our methodology of extracting the proposed feature. Finally, in Section \ref{evaluation}, we use an Extra Tree classifier to compare the predictive power of our proposed NetFlow based feature set, with the proprietary features sets provided with the original benchmark NIDS datasets. 
Finally, Section 5 concludes the paper.

\section{Limitations of Existing Datasets}
\label{lm}
Researchers have created engineered benchmark NIDS datasets because of the complexity in obtaining labelled realistic network traffic. A network testbed is designed to simulate the network behaviour of multiple end nodes. The artificial network environment overcomes the security and privacy issues faced by real-world networks. Besides, labelling the network flows generated by such controlled environments is more reliable than the open-world nature of realistic networks. During the experiments, benign network traffic and various attack scenarios are generated and conducted over the network testbed. In the meanwhile, the network packets are captured in their native packet capture (pcap) format and dumped onto storage devices. A set of network data features are extracted from the pcap files using appropriate tools and methods, forming network data flows. The result is a data source of labelled network flows reflecting benign and malicious network behaviour. The generated datasets are published and made publicly accessible for use in the design and evaluation phases of ML-based NIDS models \cite{RING2019147}.

The network data features that form these data flows are critical as they need to represent an adequate amount of security events that would aid in the ML model's classification of benign and attack classes. They also need to be feasible in count and extraction's complexity for scalable and practical deployments. A key task of designing an ML-based NIDS is the selection of the utilised data features. However, due to the lack of a standard feature set in generating NIDS datasets, the authors have applied their domain experience in the selection of these features. As a result, each available dataset is made up of its own unique set of features that their authors believe would lead to the best possible results in the classification stage. Each of the current feature sets is almost exclusive and completely different from other sets, sharing only a small number of features. The current evaluation method of ML models across multiple datasets requires the usage of the unique feature sets presented by each dataset.

The differences in the security information represented by each dataset's feature set have caused limitations and concerns regarding the reliability of the evaluation methods followed. The three main issues of not having a standard feature set are; 1. Complex extraction of several features from network traffic, some of which are irrelevant due to the lack of security events and 2. Limited ability to evaluate an ML model's generalisation to a targeted feature set across multiple datasets and 3. Lack of a universal dataset containing network data flows collected over multiple network environments. It is believed that the lack of reliable evaluation methods has caused a gap between the extensive academic research produced and the practical deployment of ML-based NIDS models in production networks \cite{sommer_paxson_2010}. Four of the most recent and widely-used NIDS datasets are discussed, which represent modern behavioural network attacks due to their production time.

\begin{itemize}
\item 
\textbf{UNSW-NB15} The Cyber Range Lab of the Australian Centre for Cyber Security (ACCS) released the widely used, UNSW-NB15, dataset in 2015. The IXIA PerfectStorm tool was utilised to generate a hybrid of testbed-based benign network activities as well as synthetic attack scenarios. The tcpdump tool was implemented to capture a total of 100 GB of pcap files. Argus and Bro-IDS, now called Zeek \cite{the_zeek_network}, and twelve additional SQL algorithms were used to extract the dataset's original 49 features \cite{moustafa-slay-2015}. The dataset contains 2,218,761 (87.35\%) benign flows and 321,283 (12.65\%) attack ones, that is, 2,540,044 flows in total.

\item 
\textbf{BoT-IoT} The Cyber Range Lab of the Australian Centre for Cyber Security (ACCS) designed a network environment in 2018 that consists of normal and botnet traffic \cite{DBLP:journals/corr/abs-1811-00701}. The Ostinato and Node-red tools were utilised to generate the non-IoT and IoT traffic respectively. A total of 69.3GB of pcap files were captured and the Argus tool was used to extract the dataset's original 42 features. The dataset contains 477 (0.01\%) benign flows and 3,668,045 (99.99\%) attack ones, that is, 3,668,522 flows in total.

\item 
\textbf{ToN-IoT} A recent heterogeneous dataset released in 2019 \cite{fesz-dm97-19} that includes telemetry data of Internet of Things (IoT) services, network traffic of IoT networks, and operating system logs. In this paper, the portion containing network traffic flows is utilised. The dataset is made up of a large number of attack scenarios conducted in a representation of a realistic large-scale network at the Cyber Range Lab by ACCS. Bro-IDS, now called Zeek \cite{the_zeek_network}, was used to extract the dataset's original 44 features. The dataset is made up of 796,380 (3.56\%) benign flows and 21,542,641 (96.44\%) attack samples, that is, 22,339,021 flows in total.

\item 
\textbf{CSE-CIC-IDS2018} A dataset released by a collaborative project between the Communications Security Establishment (CSE) \& Canadian Institute for Cybersecurity (CIC) in 2018 \cite{sharafaldin-habibi-lashkari-ghorbani-2018}. The victim network consisted of five different organisational departments and an additional server room. The benign packets were generated by network events using the abstract behaviour of human users. The attack scenarios were executed by one or more machines outside the target network. The CICFlowMeter-V3 tool was used to extract the original dataset's 75 features. The full dataset has 13,484,708 (83.07\%) benign flows and 2,748,235 (16.93\%) attack flows, that is, 16,232,943 flows in total.
\end{itemize}

\begin{figure}[!t]
    \centering
    \includegraphics[width=10cm, height=8cm]{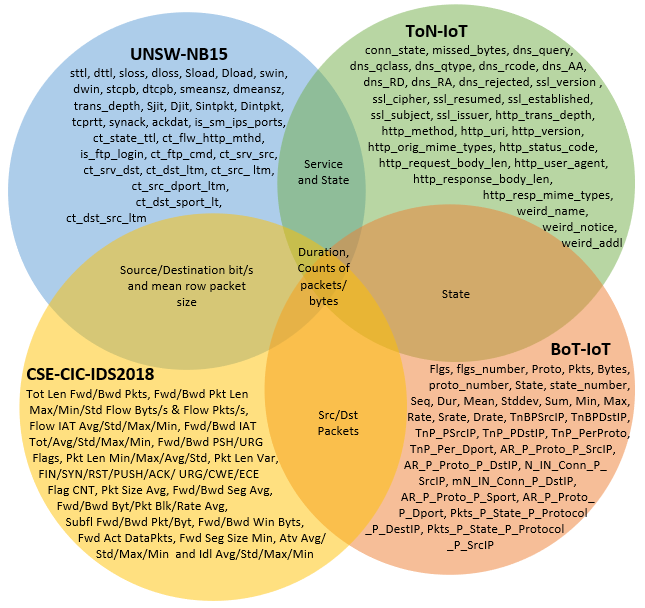}
    \caption{Venn diagram of the shared and exclusive features of four NIDS datasets
    }
    \label{SF}
\end{figure}

In Figure \ref{SF}, the shared and unique features of the aforementioned datasets are displayed. The set of features available in all four datasets contains 3 features, and the pairwise shared feature numbers vary from 1 to 5. As most of the features are exclusive to individual datasets, the evaluation of proposed ML models using a targeted feature set across the four datasets is challenging. Moreover, the ratio of the classes, i.e., benign and attack flows, is extremely varied in each dataset. Where the UNSW-NB15 and CSE-CIC-IDS2018 datasets have very high benign-to-attack ratios, whereas the ToN-IoT and BoT-IoT datasets are mainly made up of attack samples, which do not represent a realistic network behaviour. Also, some of the features in the UNSW-NB15, BoT-IoT, and CSE-CIC-IDS2018 datasets are handcrafted features that are not originally found in network packets but are statistically calculated based on other features, such as the total number of bytes transferred over the last 100 seconds. All these differences in the security information presented by the datasets have led to the design of a standard feature set for NIDS datasets.

\section{Benchmarking a Standard Feature Set}
\label{NF}
Due to the aforementioned limitations faced by current NIDS datasets made up of unique feature sets, in this paper, a standard feature set is proposed. The feature set will be evaluated and benchmarked to be used in the releases of new NIDS datasets to efficiently design ML-based NIDS. The design of ML-based NIDS requires a feature set to be extracted and scanned for intrusions when implemented. The choice of these features significantly alters the performance of the NIDS as they need to contain an adequate amount of security events to aid the ML model classification. By having a standard feature set, researchers can evaluate their model's classification ability based on their chosen features, across multiple datasets and hence different attack scenarios conducted over several network environments. This can be used to make sure their measured model performance generalises when deployed over different networks. Moreover, by having datasets sharing a common ground feature set, they can be merged to create a universal comprehensive source of data. Finally, having a standard feature set will grant control over the security information presented by NIDS datasets. We believe that a standard feature set will narrow the gap between the number of research experiments and the practical deployment of ML-based NIDS \cite{sommer_paxson_2010}.

\subsection{NetFlow}
The collection and storage of network traffic are important for organisations to monitor, analyse, and audit their network environments. However, network traffic tends to overload in volume and therefore are aggregated in terms of flows. A network data flow is a sequence of packets, in either uni- or bi-direction, between two unique endpoints sharing some attributes such as source/destination IP address and L4 (transport layer) ports, and the L4 protocol, also known as the five-tuple \cite{sarhan2020netflow}. A data flow can also be enhanced with additional features, each representing details of the respective network traffic. The information provided by these features contains security events that are essential in analysing network traffic in case of a threat \cite{Li2013}. Network flows can be represented in various formats where the NetFlow is the de-facto industry standard, developed and proposed by Darren and Barry Bruins from Cisco in 1996 \cite{Kerr2001}. NetFlow evolved over the years, where version 9 is the most common due to its larger variety of data features and bidirectional flow support \cite{CiscoSystems2011}.

\begin{table}[!h]\scriptsize
\centering
\caption{List of the proposed standard NetFlow features}
\label{nf}
\begin{tabular}{|l|l|}
\hline
\multicolumn{1}{|c|}{\textbf{Feature}} & \multicolumn{1}{c|}{\textbf{Description}} \\ \hline
IPV4\_SRC\_ADDR                        & IPv4 source address                       \\ \hline
IPV4\_DST\_ADDR                        & IPv4 destination address                  \\ \hline
L4\_SRC\_PORT                          & IPv4 source port number                   \\ \hline
L4\_DST\_PORT                          & IPv4 destination port number              \\ \hline
PROTOCOL                               & IP protocol identifier byte               \\ \hline
L7\_PROTO                              & Layer 7 protocol (numeric)                 \\ \hline
IN\_BYTES                              & Incoming number of bytes                  \\ \hline
OUT\_BYTES                             & Outgoing number of bytes                  \\ \hline
IN\_PKTS                               & Incoming number of packets                \\ \hline
OUT\_PKTS                              & Outgoing number of packets                \\
\hline
FLOW\_DURATION\_MILLISECONDS           & Flow duration in milliseconds             \\ \hline
TCP\_FLAGS                             & Cumulative of all TCP flags               \\ \hline
CLIENT\_TCP\_FLAGS                              & Cumulative of all client TCP flags                  \\ \hline
SERVER\_TCP\_FLAGS                            & Cumulative of all server TCP flags                  \\ \hline

DURATION\_IN                        & Client to Server stream duration (msec)                      \\ \hline
DURATION\_OUT                      & Client to Server stream duration (msec)                 \\ \hline
MIN\_TTL                          & Min flow TTL                   \\ \hline
MAX\_TTL                        & Max flow TTL \\ \hline
LONGEST\_FLOW\_PKT                               & Longest packet (bytes) of the flow               \\ \hline
SHORTEST\_FLOW\_PKT                            & Shortest packet (bytes) of the flow              \\ \hline
MIN\_IP\_PKT\_LEN                             & Len of the smallest flow IP packet observed                \\ \hline
MAX\_IP\_PKT\_LEN          & Len of the largest flow IP packet observed             \\ \hline
SRC\_TO\_DST\_SECOND\_BYTES                             & Src to dst Bytes/sec              \\ \hline
DST\_TO\_SRC\_SECOND\_BYTES          & Dst to src Bytes/sec        \\ \hline
RETRANSMITTED\_IN\_BYTES                        & Number of retransmitted TCP flow bytes (src-$>$dst)                       \\ \hline
RETRANSMITTED\_IN\_PKTS                     & Number of retransmitted TCP flow packets (src-$>$dst)                  \\ \hline
RETRANSMITTED\_OUT\_BYTES                          & Number of retransmitted TCP flow bytes (dst-$>$src)                   \\ \hline
RETRANSMITTED\_OUT\_PKTS                       & Number of retransmitted TCP flow packets (dst-$>$src)              \\ \hline
SRC\_TO\_DST\_AVG\_THROUGHPUT                               & Src to dst average thpt (bps)               \\ \hline
DST\_TO\_SRC\_AVG\_THROUGHPUT                           & Dst to src average thpt (bps)               \\ \hline
NUM\_PKTS\_UP\_TO\_128\_BYTES                             & Packets whose IP size $<$= 128                \\ \hline
NUM\_PKTS\_128\_TO\_256\_BYTES          & Packets whose IP size $>$ 128 and $<$= 256             \\ \hline
NUM\_PKTS\_256\_TO\_512\_BYTES                        & Packets whose IP size $>$ 256 and $<$= 512                       \\ \hline
NUM\_PKTS\_512\_TO\_1024\_BYTES                     & Packets whose IP size $>$ 512 and $<$= 1024                  \\ \hline
NUM\_PKTS\_1024\_TO\_1514\_BYTES                          & Packets whose IP size $>$  1024 and $<$= 1514                   \\ \hline
TCP\_WIN\_MAX\_IN                       &  Max TCP Window (src-$>$dst)              \\ \hline
TCP\_WIN\_MAX\_OUT                               & Max TCP Window (dst-$>$src)               \\ \hline
ICMP\_TYPE                           & ICMP Type * 256 + ICMP code               \\ \hline
ICMP\_IPV4\_TYPE                             & ICMP Type               \\ \hline
DNS\_QUERY\_ID          & DNS query transaction Id           \\ \hline
DNS\_QUERY\_TYPE                           & DNS query type (e.g., 1=A, 2=NS..)              \\ \hline
DNS\_TTL\_ANSWER                             & TTL of the first A record (if any)               \\ \hline
FTP\_COMMAND\_RET\_CODE          & FTP client command return code         \\ \hline
\end{tabular}
\end{table}

Most of the network devices such as routers and switches are capable of extracting NetFlow records. This is a great motivation for standardising NetFlow features for NIDS datasets, as the level of complexity and resources required to collect and store them is lower. In this paper, NetFlow v9 features have been utilised to form the proposed feature set, listed and described in Table \ref{nf}. There are 43 features in total with some providing information on general flow statistics and others on specific protocol applications such as DNS and FTP. All features are flow-based, meaning they are extracted from packet headers and do not depend on the payload information which is often encrypted in secure communications due to privacy concerns. The chosen features are numerical in type for efficient ML experiments. These features contain useful security events to enhance the models' intrusions detection capabilities. 


\subsection{Datasets}

\begin{figure}[!h]
    \centering
    \includegraphics[width=7cm, height=3cm]{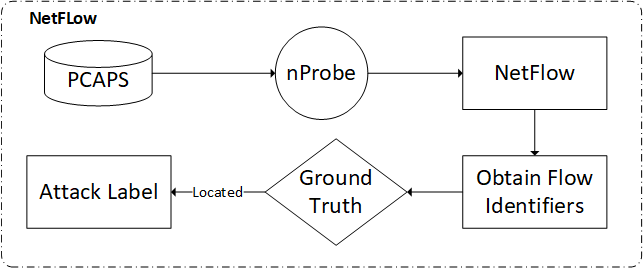}
    \caption{Feature set extraction and labelling procedure}
    \label{fig:NetFlow}
\end{figure}

Figure \ref{fig:NetFlow} shows the procedure of generating NIDS datasets using the proposed feature set. The nProbe tool by Ntop \cite{Ntopng2017} is utilised to extract 43 NetFlow version 9 features from the publicly available pcap files. The output format is chosen as text flows, in which each feature is separated by a comma (,) to be utilised as CSV files. Two label features are created by matching the five flow identifiers; source/destination IPs and ports and protocol to the ground truth attack events published by the original dataset. If a data flow is located in the attack events it would be labelled as an attack (class 1) in the binary label and its respective attack's type would be recorded in the attack label, otherwise, the sample is labelled as a benign flow (class 0).

\begin{table}[!h]\scriptsize
\centering
\caption{Specifications of the datasets proposed in this paper, compared to the original and basic NetFlow datasets}
\label{ds}
\begin{tabular}{|l|l|l|l|l|l|}
\hline
\multicolumn{1}{|c|}{\textbf{Dataset}} &
  \multicolumn{1}{c|}{\textbf{\begin{tabular}[c]{@{}c@{}}Release \\ year\end{tabular}}} &
  \multicolumn{1}{c|}{\textbf{Feature extraction tool}} &
  \multicolumn{1}{c|}{\textbf{\begin{tabular}[c]{@{}c@{}}Number \\ of features\end{tabular}}} &
  \multicolumn{1}{c|}{\textbf{\begin{tabular}[c]{@{}c@{}}CSV size\\ (GB)\end{tabular}}} &
  \multicolumn{1}{c|}{\textbf{\begin{tabular}[c]{@{}c@{}}Benign to attack \\ samples ratio\end{tabular}}} \\ \hline
UNSW-NB15          & 2015 & Argus, Bro-IDS and MS SQL & 49 & 0.55 & 8.7 to 1.3 \\
NF-UNSW-NB15      & 2020 & nProbe                   & 12 & 0.11 & 9.6 to 0.4 \\
\textbf{NF-UNSW-NB15-v2}       & \textbf{2021} & \textbf{nProbe}                    & \textbf{43} & \textbf{0.41} & \textbf{9.6 to 0.4} \\ \hline
BoT-IoT             & 2018 & Argus                   & 42 & 0.95 & 0.0 to 10 \\
NF-BoT-IoT         & 2020 & nProbe                    & 12 & 0.05 & 0.2 to 9.8 \\
\textbf{NF-BoT-IoT-v2}          & \textbf{2021} & \textbf{nProbe}                    & \textbf{43} & \textbf{5.60} & \textbf{0.0 to 10.0} \\ \hline
ToN-IoT            & 2020 & Bro-IDS                   & 44 & 3.02 & 0.4 to 9.6 \\
NF-ToN-IoT         & 2020 & nProbe                    & 12 & 0.09 & 2.0 to 8.0\\
\textbf{NF-ToN-IoT-v2}         & \textbf{2021} & \textbf{nProbe}                    & \textbf{43} & \textbf{2.47} & \textbf{3.6 to 6.4} \\ \hline
CSE-CIC-IDS2018    & 2018 & CICFlowMeter-V3           & 75 & 6.41 & 8.3 to 1.7 \\
NF-CSE-CIC-IDS2018 & 2020 & nProbe                    & 12 & 0.58 & 8.8 to 1.2 \\
\textbf{NF-CSE-CIC-IDS2018-v2} & \textbf{2021} & \textbf{nProbe}                    & \textbf{43} & \textbf{2.80} & \textbf{8.8 to 1.2} \\ \hline
NF-UQ-NIDS         & 2020 & nProbe                   & 12 & 1.0 & 7.7 to 2.3 \\
\textbf{NF-UQ-NIDS-v2}         & \textbf{2021} & \textbf{nProbe}                    & \textbf{43} & \textbf{12.5} & \textbf{3.3 to 6.7} \\ \hline
\end{tabular}%
\end{table}

In this paper, the proposed feature set has been extracted from four well-known datasets; UNSW-NB15, BoT-IoT, ToN-IoT, and CSE-CIC-IDS2018. Their publicly available pcap files and ground truth events have been utilised in the features extraction and labelling processes respectively. The generated datasets have been named NF-UNSW-NB15-v2, NF-BoT-IoT-v2, NF-ToN-IoT-v2, NF-CSE-CIC-IDS2018-v2 and NF-UQ-NIDS-v2. The later dataset is a merge of all other datasets, which is a practical advantage of having a common feature set. Table \ref{ds} lists the NetFlow datasets and compares their properties to the original datasets in terms of the Feature Extraction (FE) tool utilised, the number of features, file size and the benign to attack samples ratio. As illustrated, two NetFlow datasets are corresponding to each original NIDS dataset, where v1 and v2 are the basic and extended versions respectively. The fifth NetFlow dataset is a comprehensive dataset that combines all four.

\begin{itemize}
\item 
\textbf{NF-UNSW-NB15-v2} The NetFlow-based format of the UNSW-NB15 dataset, named NF-UNSW-NB15, has been extended with additional NetFlow features and labelled with its respective attack categories. The total number of data flows are 2,390,275 out of which 95,053 (3.98\%) are attack samples and 2,295,222 (96.02\%) are benign. The attack samples are further classified into nine subcategories, Table \ref{un} represents the NF-UNSW-NB15-v2 dataset's distribution of all flows.
\end{itemize}

\begin{table}[!h]\scriptsize\centering
\caption{NF-UNSW-NB15-v2 distribution}
\label{un}
\begin{tabular}{|l|l|l|}
\hline
\multicolumn{1}{|c|}{\textbf{Class}} & \multicolumn{1}{c|}{\textbf{Count}} & \multicolumn{1}{c|}{\textbf{Description}}                                                     \\ \hline
Benign                               & 2295222                             & Normal unmalicious flows                                                                      \\ \hline
Fuzzers &
  22310 &
  \begin{tabular}[c]{@{}l@{}}An attack in which the attacker sends large amounts of random data which cause a system \\ to crash and also aim to discover security vulnerabilities in a system.\end{tabular} \\ \hline
Analysis                             & 2299                                &  \begin{tabular}[c]{@{}l@{}} A group that presents a variety of threats that target web applications through ports, \\emails and scripts. \end{tabular} \\ \hline
Backdoor                             & 2169                                & \begin{tabular}[c]{@{}l@{}} A technique that aims to bypass security mechanisms by replying to specific constructed \\client applications.   \end{tabular}\\ \hline
DoS &
  5794 &
  \begin{tabular}[c]{@{}l@{}}Denial of Service is an attempt to overload a computer system’s resources with the aim \\ of preventing access to or availability of its data.\end{tabular} \\ \hline
Exploits                             & 31551                               & \begin{tabular}[c]{@{}l@{}}Are sequences of commands controlling the behaviour of a host through a known \\vulnerability. \end{tabular} \\ \hline
Generic                              & 16560                                & A method that targets cryptography and causes a collision with each block-cipher.          \\ \hline
Reconnaissance                       & 12779                               & A technique for gathering information about a network host and is also known as a probe.   \\ \hline
Shellcode                            & 1427                                & A malware that penetrates a code to control a victim’s host.                                 \\ \hline
Worms                                & 164                                 & Attacks that replicate themselves and spread to other computers.                          \\ \hline
\end{tabular}
\end{table}

\begin{itemize}
\item 
\textbf{NF-BoT-IoT-v2} An IoT NetFlow-based dataset is generated by expanding the NF-BoT-IoT dataset. The features were extracted from the publicly available pcap files and the flows were labelled with their respective attack categories. The total number of data flows are 37,763,497 out of which 37,628,460 (99.64\%) are attack samples and 135,037 (0.36\%) are benign. There are four attack categories in the dataset, Table \ref{bo} represents the NF-BoT-IoT-v2 distribution of all flows.
\end{itemize}

\begin{table}[!h]\scriptsize\centering
\caption{NF-BoT-IoT-v2 distribution}
\label{bo}
\begin{tabular}{|l|l|l|}
\hline
\multicolumn{1}{|c|}{\textbf{Class}} & \multicolumn{1}{c|}{\textbf{Count}} & \multicolumn{1}{c|}{\textbf{Description}}                                                     \\ \hline
Benign                               & 135037                             & Normal unmalicious flows                                                                      \\ \hline
Reconnaissance &
  2620999 &
  A technique for gathering information about a network host and is also known as a probe. \\ \hline
DDoS                             & 18331847                                &  \begin{tabular}[c]{@{}l@{}}Distributed Denial of Service is an attempt similar to DoS but has multiple \\ different distributed sources.\end{tabular} \\ \hline
DoS                             & 16673183                                &  \begin{tabular}[c]{@{}l@{}}An attempt to overload a computer system’s resources with the aim of preventing access\\ to or availability of its data.\end{tabular}\\ \hline
Theft &
  2431 &
  \begin{tabular}[c]{@{}l@{}}A group of attacks that aims to obtain sensitive data such as data theft and keylogging\end{tabular} \\ \hline
\end{tabular}
\end{table}

\begin{itemize}
\item 
\textbf{NF-ToN-IoT-v2} The publicly available pcaps of the ToN-IoT dataset are utilised to generate its NetFlow records, leading to a NetFlow-based IoT network dataset called NF-ToN-IoT. The total number of data flows are 16,940,496 out of which 10,841,027 (63.99\%) are attack samples and 6,099,469 (36.01\%) are benign. Table \ref{ton} lists and defines the distribution of the NF-ToN-IoT-v2 dataset.
\end{itemize}

\begin{table}[!h]\scriptsize\centering
\caption{NF-ToN-IoT-v2 distribution}
\label{ton}
\begin{tabular}{|l|l|l|}
\hline
\multicolumn{1}{|c|}{\textbf{Class}} &
  \multicolumn{1}{c|}{\textbf{Count}} &
  \multicolumn{1}{c|}{\textbf{Description}} \\ \hline
Benign &
  6099469 &
  Normal unmalicious flows \\ \hline
Backdoor &
  16809 &
  \begin{tabular}[c]{@{}l@{}}A technique that aims to attack remote-access computers by replying to specific constructed \\ client applications \end{tabular} \\ \hline
DoS &
  712609 &
  \begin{tabular}[c]{@{}l@{}}An attempt to overload a computer system’s resources with the aim of preventing access to or\\ availability of its data.\end{tabular} \\ \hline
DDoS &
  2026234 &
  \begin{tabular}[c]{@{}l@{}}An attempt similar to DoS but has multiple \\ different distributed sources.\end{tabular} \\ \hline
Injection &
  684465 &
  \begin{tabular}[c]{@{}l@{}}A variety of attacks that supply untrusted inputs that aim to alter the course of \\ execution, with SQL and Code injections two of the main ones.\end{tabular} \\ \hline
MITM &
  7723 &
  \begin{tabular}[c]{@{}l@{}} Man In The Middle is a method that places an attacker between a victim and host with which \\ the victim is trying to communicate, with the aim of intercepting traffic and communications.\end{tabular} \\ \hline
Password &
  1153323 &
  covers a variety of attacks aimed at retrieving passwords by either brute force or sniffing. \\ \hline
Ransomware &
  3425 &
  \begin{tabular}[c]{@{}l@{}} An attack that encrypts the files stored on a host and asks for compensation in exchange for \\ the decryption technique/key.\end{tabular} \\ \hline
Scanning &
  3781419 &
  \begin{tabular}[c]{@{}l@{}}A group that consists of a variety of techniques that aim to discover information about networks \\ and hosts, and is also known as probing.\end{tabular} \\ \hline
XSS &
  2455020 &
  \begin{tabular}[c]{@{}l@{}}Cross-site Scripting is a type of injection in which an attacker uses web applications to send \\ malicious scripts to end-users.\end{tabular} \\ \hline
\end{tabular}
\end{table}

\begin{itemize}
\item
\textbf{NF-CSE-CIC-IDS2018-v2} The original pcap files of the CSE-CIC-IDS2018 dataset are utilised to generate a NetFlow-based dataset called NF-CSE-CIC-IDS2018-v2. The total number of flows are 18,893,708 out of which 2,258,141 (11.95\%) are attack samples and 16,635,567 (88.05\%) are benign ones, Table \ref{cse} represents the dataset's distribution.
\end{itemize}

\begin{table}[!h]\scriptsize\centering
\caption{NF-CSE-CIC-IDS2018-v2 distribution}
\label{cse}
\begin{tabular}{|l|l|l|}
\hline
\multicolumn{1}{|c|}{\textbf{Class}} &
  \multicolumn{1}{c|}{\textbf{Count}} &
  \multicolumn{1}{c|}{\textbf{Description}} \\ \hline
Benign &
  16635567 &
  Normal unmalicious flows \\ \hline
BruteForce &
  120912 &
   \begin{tabular}[c]{@{}l@{}}A technique that aims to obtain usernames and password credentials by accessing a list of \\predefined possibilities \end{tabular}\\ \hline
Bot &
  143097 &
  \begin{tabular}[c]{@{}l@{}}An attack that enables an attacker to remotely control several hijacked computers to perform\\ malicious activities.\end{tabular} \\ \hline
DoS &
  483999 &
  \begin{tabular}[c]{@{}l@{}}An attempt to overload a computer system’s resources with the aim of preventing access to or \\ availability of its data.\end{tabular} \\ \hline
DDoS &
  1390270 &
  An attempt similar to DoS but has multiple different distributed sources. \\ \hline
Infiltration &
  116361 &
  \begin{tabular}[c]{@{}l@{}}An inside attack that sends a malicious file via an email to exploit an application and is \\followed by a backdoor that scans the network for other vulnerabilities\end{tabular} \\ \hline
Web Attacks &
  3502 &
  A group that includes SQL injections, command injections and unrestricted file uploads \\ \hline
\end{tabular}
\end{table}

\begin{itemize}
\item  
\textbf{NF-UQ-NIDS-v2} A comprehensive dataset, merging all the aforementioned datasets. The newly published dataset represents the benefits of the shared dataset feature sets, where the merging of multiple smaller datasets is possible. This will eventually lead to a bigger and universal NIDS dataset containing flows from multiple network setups and different attack settings. It includes an additional label feature, identifying the original dataset of each flow. This can be used to compare the same attack scenarios conducted over two or more different testbed networks. The attack categories have been modified to combine all parent categories. Attacks named DoS attacks-Hulk, DoS attacks-SlowHTTPTest, DoS attacks-GoldenEye and DoS attacks-Slowloris have been renamed to the parent DoS category. Attacks named DDoS attack-LOIC-UDP, DDoS attack-HOIC and DDoS attacks-LOIC-HTTP have been renamed to DDoS. Attacks named FTP-BruteForce, SSH-Bruteforce, Brute Force -Web and Brute Force -XSS have been combined as a brute-force category. Finally, SQL Injection attacks have been included in the injection attacks category. The NF-UQ-NIDS dataset has a total of 75,987,976 records, out of which 25,165,295 (33.12\%) are benign flows and 50,822,681 (66.88\%) are attacks. Table \ref{uq} lists the distribution of the final attack categories.
\end{itemize}

\begin{table}[!h]\scriptsize
\centering
\caption{NF-UQ-NIDS-v2 distribution}
\label{uq}
\begin{tabular}{||l|l||l|l||}
\hline

\textbf{Class} & \textbf{Count} &\textbf{Class} & \textbf{Count} \\ \hline \hline
Benign         & 25165295         &Scanning       & 3781419          \\ \hline
DDoS           & 21748351         &Fuzzers        & 22310          \\ \hline
Reconnaissance & 2633778         &Backdoor       & 18978          \\ \hline
Injection      & 684897         &Bot            & 143097          \\ \hline
DoS            & 17875585         &Generic        & 16560           \\ \hline
Brute Force    & 123982         &Analysis       & 2299           \\ \hline
Password       & 1153323         &Shellcode      & 1427           \\ \hline
XSS            & 2455020          &MITM           & 7723           \\ \hline
Infilteration  & 116361          &Worms          & 164            \\ \hline
Exploits       & 31551         &Ransomware     & 3425            \\ \hline
Theft       & 2431 &&\\ \hline
\end{tabular}
\end{table} 

\section{Evaluation}
\label{evaluation}

In this section, the proposed NetFlow feature set is evaluated across five NIDS datasets; NF-UNSW-NB15-v2, NF-BoT-IoT-v2, NF-ToN-IoT-v2, NF-CSE-CIC-IDS2018-v2 and NF-UQ-NIDS-v2. An ensemble ML classifier, known as Extra Trees, that belongs to the \textit{trees} family has been utilised for this purpose. The evaluation is conducted by comparing the classifier performance with the corresponding metrics of the basic NetFlow and original datasets. Various classification metrics are collected such as \textit{accuracy}, \textit{Area Under the Curve (AUC)}, \textit{F1 Score}, \textit{Detection Rate (DR)}, \textit{False Alarm Rate (FAR)} and time required to predict a single test sample in microseconds (\textmu s). As part of the data pre-processing, the flow identifiers such as IDs, source/destination IP and ports, timestamps, and start/end time are dropped to avoid learning bias towards attacking and victim end nodes. For the UNSW-NB15 and NF-UNSW-NB15-v2 datasets, The Time To Live (TTL)-based features are dropped due to their extreme correlation with the labels. Additionally, the min-max normalisation technique has been applied to scale all datasets' values between 0 and 1.  The datasets have been split into 70\%-30\% for training and testing purposes. For a fair evaluation, five cross-validation splits are conducted and the mean is measured.

\subsection{Binary-class Classification}

In Table \ref{m1}, the attack detection (binary classification) performance of the datasets has been measured and compared to the original and basic NetFlow datasets. Using the NF-UNSW-NB15-v2 dataset, the ML model's performance has significantly increased with an AUC of 0.9845, compared to 0.9485 and 0.9545 when using the NF-UNSW-NB15 and UNSW-NB15 datasets respectively. The model achieved the highest F1 score of 0.97 in the shortest prediction time when using the extended NetFlow feature set. The NF-BoT-IoT-v2 dataset has enabled the ML model to achieve the highest possible detection accuracy and F1 score, the same as the BoT-IoT dataset. However, the model has a significantly lower FAR and prediction time, resulting in an increased AUC of 0.9987 and a lower prediction time of 3.90 \textmu s compared. Using the extended NetFlow feature set, the ML model achieved significantly higher accuracy than NF-BoT-IoT of 100\% compared to 93.82\%.

\begin{table}[!h]\scriptsize
\centering
\caption{Binary-class classification results}
\label{m1}

\begin{tabular}{|l|r|r|r|r|r|r|}
\hline
\textbf{Dataset} &
  \multicolumn{1}{c|}{\textbf{Accuracy}} &
  \multicolumn{1}{c|}{\textbf{AUC}} &
  \multicolumn{1}{c|}{\textbf{F1 Score}} &
  \multicolumn{1}{c|}{\textbf{DR}} &
  \multicolumn{1}{c|}{\textbf{FAR}} &
  \multicolumn{1}{c|}{\textbf{Prediction Time (\textmu s)}} \\ \hline
UNSW-NB15         & 99.25\% & 0.9545 & 0.92 & 91.25\% & 0.35\% & 10.05 \\
NF-UNSW-NB15       & 98.62\% & 0.9485 & 0.85 & 90.70\% & 1.01\% & 7.79  \\
\textbf{NF-UNSW-NB15-v2}       & \textbf{99.73\%} & \textbf{0.9845} & \textbf{0.97} & \textbf{97.07\%} & \textbf{0.16\%} & \textbf{5.92}  \\ \hline
BoT-IoT         & 100.00\% & 0.9948 & 1.00 & 100.00\% & 1.05\% & 7.62 \\
NF-BoT-IoT       & 93.82\% & 0.9628 & 0.97 & 93.70\% & 1.13\% & 5.37  \\
\textbf{NF-BoT-IoT-v2}       & \textbf{100.00\%} & \textbf{0.9987} & \textbf{1.00} & \textbf{100.00\%} & \textbf{0.26\%} & \textbf{3.90}  \\ \hline
ToN-IoT            & 97.86\% & 0.9788 & 0.99 & 97.86\% & 2.10\% & 8.93  \\
NF-ToN-IoT         & 99.66\% & 0.9965 & 1.00 & 99.67\% & 0.37\% & 6.05\\
\textbf{NF-ToN-IoT-v2}         & \textbf{99.64\%} & \textbf{0.9959} & \textbf{1.00} & \textbf{99.76\%} & \textbf{0.58\%} & \textbf{8.47}  \\ \hline
CSE-CIC-IDS2018    & 98.31\% & 0.9684 & 0.94 & 94.75\% & 1.07\% & 23.01 \\
NF-CSE-CIC-IDS2018 & 95.33\% & 0.9506 & 0.83 & 94.71\% & 4.59\% & 17.04 \\
\textbf{NF-CSE-CIC-IDS2018-v2} & \textbf{99.35\%} & \textbf{0.9829} & \textbf{0.97} & \textbf{96.89\%} & \textbf{0.31\%} & \textbf{21.75} \\ \hline
NF-UQ-NIDS        & 97.25\% & 0.9669 & 0.94 & 95.66\% & 2.27\% & 14.35\\
\textbf{NF-UQ-NIDS-v2}        & \textbf{97.90\%} & \textbf{0.9830} & \textbf{0.98} & \textbf{97.12\%} & \textbf{0.52\%} & \textbf{14.18} \\ \hline
\end{tabular}%
\end{table}

The intrusion detection results of the ML model using the NF-ToN-IoT-v2 dataset are superior to its original ToN-IoT dataset. Compared to NF-ToN-IoT, it achieved a higher DR of but a slightly higher FAR. Overall, the accuracy achieved by the model using the NF-ToN-IoT-v2 is 99.64\%, which is higher than ToN-IoT (97.86\%) and similar to NF-ToN-IoT (99.66\%). The model performance when using the NF-CSE-CIC-IDS2018-v2 dataset is notably more efficient than the CSE-CIC-IDS2018 and NF-CSE-CIC-IDS2018-v2 datasets. It achieved a high DR of 96.89\% and a low FAR of 0.31\% and required 21.75 \textmu s per sample prediction. The overall accuracy achieved is 99.35\%, which is higher than both the CSE-CIC-IDS2018 (98.31\%) and NF-CSE-CIC-IDS2018 (95.33\%) datasets. The merged NF-UQ-NIDS-v2 dataset enabled the model to achieve an accuracy of 97.90\%, a DR of 97.12\% and a FAR of 0.52\%, outperforming the NF-UQ-NIDS dataset with a lower prediction time of 14.18 \textmu s.

\begin{figure}[!h]
    \centering
    \includegraphics[width=10cm, height=4.5cm]{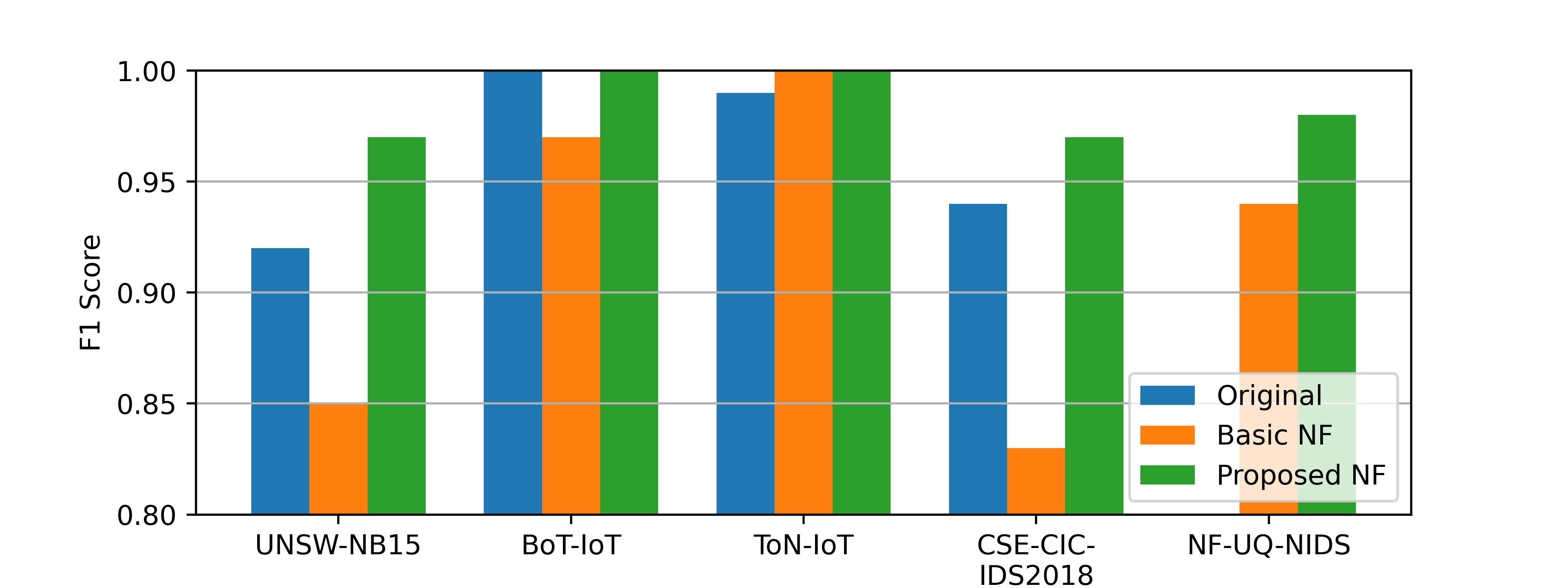}
    \caption{Binary-class classification F1 score}
    \label{we}
\end{figure}

Figure \ref{we} visually represents the F1 score obtained when applying an \textit{Extra Trees} classifier on the three different feature sets of five NIDS datasets; the original as well as basic and proposed NetFlow feature sets. This fair comparison between the NetFlow feature sets demonstrates the benefit of having a common feature set across multiple datasets. It enables the evaluation of various attack detections using a common feature set. Overall, the proposed (extended) NetFlow feature set has outperformed the original and basic feature sets in terms of attack detection performance. All datasets have significantly achieved a higher or similar F1 score to their respective datasets. It is clear that using the proposed feature set achieves a reliable detection performance. Further feature selection experiments are required to identify its key features to enhance the extraction tasks.

\subsection{Multi-class Classification}

 \begin{table}[!h]\scriptsize
\centering
\caption{NF-UNSW-NB15-v2 multi-class classification results}
\label{tab:my-table1}
\begin{tabular}{l|r|r|r|r|r|r|}
\cline{2-7}
                                               & \multicolumn{2}{c|}{\textbf{UNSW-NB15}} & \multicolumn{2}{c|}{\textbf{NF-UNSW-NB15}} & \multicolumn{2}{c|}{\textbf{NF-UNSW-NB15-v2}} \\ \hline
\multicolumn{1}{|c|}{\textbf{Class Name}} &
  \multicolumn{1}{c|}{\textbf{DR}} &
  \multicolumn{1}{c|}{\textbf{F1 Score}} &
  \multicolumn{1}{c|}{\textbf{DR}} &
  \multicolumn{1}{c|}{\textbf{F1 Score}}&
  \multicolumn{1}{c|}{\textbf{DR}} &
  \multicolumn{1}{c|}{\textbf{F1 Score}} \\ \hline
\multicolumn{1}{|l|}{{Benign}}          & 99.72\%              & 1.00             & 99.02\%               & 0.99   & 99.85\%               & 1.00            \\ \hline

\multicolumn{1}{|l|}{{Analysis}}        & 4.39\%               & 0.03             & 28.28\%               & 0.15   & 30.89\%               & 0.17             \\ \hline

\multicolumn{1}{|l|}{{Backdoor}}        & 13.96\%              & 0.08             & 39.17\%               & 0.17    & 40.30\%               & 0.18            \\ \hline

\multicolumn{1}{|l|}{{DoS}}             & 13.63\%              & 0.18             & 31.84\%               & 0.41   & 29.57\%               & 0.36             \\ \hline

\multicolumn{1}{|l|}{{Exploits}}        & 83.25\%              & 0.80             & 81.04\%               & 0.82  & 80.41\%               & 0.84             \\ \hline

\multicolumn{1}{|l|}{{Fuzzers}}         & 50.50\%              & 0.57             & 62.63\%               & 0.55   & 80.57\%               & 0.85             \\ \hline

\multicolumn{1}{|l|}{{Generic}}         & 86.08\%              & 0.91             & 57.13\%               & 0.66  & 85.15\%               & 0.90              \\ \hline
\multicolumn{1}{|l|}{{Reconnaissance}}  & 75.90\%              & 0.80             & 76.89\%               & 0.82   & 80.02\%               & 0.83             \\ \hline
\multicolumn{1}{|l|}{{Shellcode}}       & 53.61\%              & 0.59             & 87.91\%               & 0.75   & 87.67\%               & 0.69             \\ \hline
\multicolumn{1}{|l|}{{Worms}}           & 5.26\%               & 0.09             & 52.91\%               & 0.55    & 85.98\%               & 0.69           \\ \hline
\multicolumn{1}{|l|}{\textbf{Weighted Average}}         & \textbf{98.19\%}              & \textbf{0.98}             & \textbf{97.62\%}               & \textbf{0.98}     & \textbf{98.90\%}               & \textbf{0.99}            \\ \hline
\multicolumn{1}{|l|}{\textbf{Prediction Time (\textmu s)}} & \multicolumn{2}{c|}{\textbf{9.94}}      & \multicolumn{2}{c|}{\textbf{9.35}}      & \multicolumn{2}{c|}{\textbf{8.81}}     \\ \hline
\end{tabular}
\end{table}

To further evaluate the proposed NetFlow feature set, multi-classification experiments are conducted to measure the weighted average of DR, F1 score and prediction time of each class present in the datasets. Tables \ref{tab:my-table1}, \ref{bot} \ref{tab:my-table}, \ref{tab:my-table3} and \ref{tab:uq} represent the performances of the NF-UNSW-NB15-v2, NF-BoT-IoT-v2, NF-ToN-IoT-v2, NF-CSE-CIC-IDS201-v2 and NF-UQ-NIDS-v2 datasets respectively. The datasets made up of the original and basic NetFlow feature sets are provided for comparison purposes. In Table \ref{tab:my-table1}, the benefits of using the NF-UNSW-NB15-v2 over the former datasets are realised by increasing the Ml model's F1 score to 0.99 from 0.98 and decreasing the prediction time to 8.81 \textmu s. The DR of certain attack types such as fuzzers, generic, and worms have significantly improved while the others have remained at slightly the same rate. The detection of the analysis, backdoor and DoS attacks are still unreliable when using the extended Netflow feature set, further analysis is required to identify the missing key features. However, due to their small number of samples, the overall accuracy of the NF-UNSW-NB15-v2 is higher (98.90\%) than UNSW-NB15 (98.19\%) and NF-UNSW-NB15 (97.62\%).

\begin{table}[!h]\scriptsize
\centering
\caption{NF-BoT-IoT-v2 multi-class classification results}
\label{bot}
\begin{tabular}{l|r|r|r|r|r|r|}
\cline{2-7}
\textbf{}                                     & \multicolumn{2}{c|}{\textbf{BoT-IoT}} & \multicolumn{2}{c|}{\textbf{NF-BoT-IoT}} & \multicolumn{2}{c|}{\textbf{NF-BoT-IoT-v2}}\\ \hline
\multicolumn{1}{|c|}{\textbf{Class Name}} &
  \multicolumn{1}{c|}{\textbf{DR}} &
  \multicolumn{1}{c|}{\textbf{F1 Score}} &
  \multicolumn{1}{c|}{\textbf{DR}} &
  \multicolumn{1}{c|}{\textbf{F1 Score}}&
  \multicolumn{1}{c|}{\textbf{DR}} &
  \multicolumn{1}{c|}{\textbf{F1 Score}} \\ \hline
\multicolumn{1}{|l|}{{Benign}}         & 99.58\%             & 0.99            & 98.65\%              & 0.43 & 99.76\%              & 1.00             \\ \hline
\multicolumn{1}{|l|}{{DDoS}}           & 100.00\%            & 1.00            & 30.37\%              & 0.28 & 99.99\%              & 1.00             \\ \hline
\multicolumn{1}{|l|}{{DoS}}            & 100.00\%            & 1.00            & 36.33\%              & 0.31   & 99.99\%              & 1.00           \\ \hline
\multicolumn{1}{|l|}{{Reconnaissance}} & 100.00\%            & 1.00            & 89.95\%              & 0.90   & 99.93\%              & 1.00           \\ \hline
\multicolumn{1}{|l|}{{Theft}}          & 91.16\%             & 0.95           & 88.06\%              & 0.18   & 83.01\%              & 0.85          \\ \hline
\multicolumn{1}{|l|}{\textbf{Weighted Average}} &
  \textbf{100.00\%} &
  \textbf{1.00} &
  \textbf{73.58\%} &
  \textbf{0.77}&
  \textbf{99.99\%} &
  \textbf{1.00} \\ \hline
\multicolumn{1}{|l|}{\textbf{Prediction Time (\textmu s)}} &
  \multicolumn{2}{c|}{\textbf{12.63}} &
  \multicolumn{2}{c|}{\textbf{9.19}}&
  \multicolumn{2}{c|}{\textbf{11.86}} \\ \hline
\end{tabular}%
\end{table}

Table \ref{bot} shows that when using the NF-BoT-IoT-v2 dataset, the ML model is achieving the almost perfect multi-classification performance gained when using the BoT-IoT dataset of 100\% accuracy and 1.00 F1 Score. The four attack categories are almost fully detected except for the theft attacks, where only 83.01\% were successfully detected. The accuracy of the ML model is increased from 73.58\% to 99.99\% and the F1 score from 0.77 to 1.00 when applied to the extended NetFlow feature set compared to the basic set. Overall, it is a significant improvement that overcomes the performance limitations faced by the basic NetFlow datasets, despite the slight increase in prediction time.

\begin{table}[!h]\scriptsize
\centering
\caption{NF-ToN-IoT-v2 multi-class classification results}
\label{tab:my-table}
\begin{tabular}{l|l|l|l|l|l|l|}
\cline{2-7}
\textbf{}                                      & \multicolumn{2}{c|}{\textbf{ToN-IoT}}       & \multicolumn{2}{c|}{\textbf{NF-ToN-IoT}} & \multicolumn{2}{c|}{\textbf{NF-ToN-IoT-v2}}   \\ \hline
\multicolumn{1}{|l|}{\textbf{Class Name}}      & \textbf{DR} & \textbf{F1 Score} & \textbf{DR} & \textbf{F1 Score}& \textbf{DR} & \textbf{F1 Score} \\ \hline
\multicolumn{1}{|l|}{{Benign}}     & 89.97\% & 0.94 & 98.97\% & 0.99 & 99.44\% & 0.99 \\ \hline
\multicolumn{1}{|l|}{{Backdoor}}   & 98.05\% & 0.31 & 99.22\% & 0.98 & 99.79\% & 1.00 \\ \hline
\multicolumn{1}{|l|}{{DDoS}}       & 96.90\% & 0.98 & 63.22\% & 0.72 & 98.76\% & 0.99 \\ \hline
\multicolumn{1}{|l|}{{DoS}}        & 53.89\% & 0.57 & 95.91\% & 0.48 & 89.41\% & 0.91 \\ \hline
\multicolumn{1}{|l|}{{Injection}}  & 96.67\% & 0.96 & 41.47\% & 0.51 & 90.14\% & 0.91 \\ \hline
\multicolumn{1}{|l|}{{MITM}}       & 66.25\% & 0.16 & 52.81\% & 0.38 & 37.45\% & 0.45 \\ \hline
\multicolumn{1}{|l|}{{Password}}   & 86.99\% & 0.92 & 27.36\% & 0.24 & 97.16\% & 0.97 \\ \hline
\multicolumn{1}{|l|}{{Ransomware}} & 89.87\% & 0.11 & 87.33\% & 0.83 & 97.29\% & 0.98 \\ \hline
\multicolumn{1}{|l|}{{Scanning}}   & 75.05\% & 0.85 & 31.30\% & 0.08 & 99.67\% & 1.00 \\ \hline
\multicolumn{1}{|l|}{{XSS}}        & 98.83\% & 0.99 & 24.49\% & 0.19 & 96.83\% & 0.96 \\ \hline
\multicolumn{1}{|l|}{\textbf{Weighted Average}}    & \textbf{84.61\%} & \textbf{0.87} & \textbf{56.34\%} & \textbf{0.60}& \textbf{98.05\%} & \textbf{0.98} \\ \hline
\multicolumn{1}{|l|}{\textbf{Prediction Time (\textmu s)}} & \multicolumn{2}{c|}{\textbf{12.02}}         & \multicolumn{2}{c|}{\textbf{21.21}}    & \multicolumn{2}{c|}{\textbf{12.15}}     \\ \hline
\end{tabular}
\end{table}

In Table \ref{tab:my-table}, the NF-ToN-IoT-v2 dataset has enabled the ML model to achieve outstanding results when conducting multi-classification experiments. The extended NetFlow feature set notably outperformed both the ToN-IoT and NF-ToN-IoT feature sets by increasing the model's weighted F1 score to 0.98 from 0.87 and 0.60 respectively. The model also requires a lower prediction time compared to when applied to the basic NetFlow dataset. The extended Netflow feature set has increased the DR of all attack types except for DoS, MITM, and XSS attacks. Further analysis of features containing useful security events is essential to aid in their detection. Overall, the feature set of NF-ToN-IoT-v2 has aided the ML model in the detection of the attacks present in the dataset, with an enhanced accuracy of 98.05\% confirming the reliability of the extended NetFlow feature set.

\begin{table}[!h]\scriptsize
\centering
\caption{NF-CSE-CIC-IDS2018-v2 multi-class classification results}
\label{tab:my-table3}
\begin{tabular}{l|l|l|l|l|l|l|}
\cline{2-7}
\textbf{}                                      & \multicolumn{2}{c|}{\textbf{CSE-CIC-IDS2018}} & \multicolumn{2}{c|}{\textbf{NF-CSE-CIC-IDS2018}}& \multicolumn{2}{c|}{\textbf{NF-CSE-CIC-IDS2018-v2}} \\ \hline
\multicolumn{1}{|l|}{\textbf{Class Name}}      & \textbf{DR}  & \textbf{F1 Score}  & \textbf{DR}    & \textbf{F1 Score} & \textbf{DR}    & \textbf{F1 Score}  \\ \hline
\multicolumn{1}{|l|}{{Benign}}                   & 89.50\%  & 0.94 & 69.83\%  & 0.82 & 99.69\%  & 1.00\\ \hline
\multicolumn{1}{|l|}{{Bot}}                      & 99.92\%  & 0.99 & 100.00\% & 1.00 & 100.00\%  & 1.00\\ \hline
\multicolumn{1}{|l|}{{Brute Force -Web}}         & 71.36\%  & 0.01 & 50.21\%  & 0.52 & 28.05\%  & 0.01\\ \hline
\multicolumn{1}{|l|}{{Brute Force -XSS}}         & 72.17\%  & 0.72 & 49.16\%  & 0.39 & 29.34\%  & 0.00\\ \hline
\multicolumn{1}{|l|}{{DDoS attack-HOIC}}         & 100.00\% & 1.00 & 45.66\%  & 0.39 & 57.33\%  & 0.73\\ \hline
\multicolumn{1}{|l|}{{DDoS attack-LOIC-UDP}}     & 83.59\%  & 0.82 & 80.98\%  & 0.82 & 99.29\%  & 1.00\\ \hline
\multicolumn{1}{|l|}{{DDoS attacks-LOIC-HTTP}}   & 99.93\%  & 1.00 & 99.93\%  & 0.71 & 100.00\%  & 1.00\\ \hline
\multicolumn{1}{|l|}{{DoS attacks-GoldenEye}}    & 99.97\%  & 1.00 & 99.32\%  & 0.98 & 100.00\%  & 1.00\\ \hline
\multicolumn{1}{|l|}{{DoS attacks-Hulk}}         & 100.00\% & 1.00 & 99.65\%  & 0.99 & 100.00\%  & 1.00\\ \hline
\multicolumn{1}{|l|}{{DoS attacks-SlowHTTPTest}} & 69.80\%  & 0.60 & 0.00\%   & 0.00 & 100.00\%  & 1.00\\ \hline
\multicolumn{1}{|l|}{{DoS attacks-Slowloris}}    & 99.44\%  & 0.62 & 99.95\%  & 1.00 & 99.99\%  & 1.00\\ \hline
\multicolumn{1}{|l|}{{FTP-BruteForce}}           & 68.76\%  & 0.75 & 100.00\% & 0.79 & 100.00\%  & 1.00\\ \hline
\multicolumn{1}{|l|}{{Infilteration}}            & 36.15\%  & 0.08 & 62.66\%  & 0.04 & 39.58\%  & 0.43\\ \hline
\multicolumn{1}{|l|}{{SQL Injection}}            & 49.34\%  & 0.30 & 25.00\%  & 0.22 & 41.44\%  & 0.00\\ \hline
\multicolumn{1}{|l|}{{SSH-Bruteforce}}           & 99.99\%  & 1.00 & 99.93\%  & 1.00 & 100.00\%  & 1.00\\ \hline
\multicolumn{1}{|l|}{\textbf{Weighted Average}}                  & \textbf{90.28\%}  & \textbf{0.94} & \textbf{71.92\%}  & \textbf{0.80} & \textbf{96.90\%}  & \textbf{0.98}\\ \hline
\multicolumn{1}{|l|}{\textbf{Prediction Time (\textmu s)}} & \multicolumn{2}{c|}{\textbf{24.17}}           & \multicolumn{2}{c|}{\textbf{17.29}}    & \multicolumn{2}{c|}{\textbf{27.28}}            \\ \hline
\end{tabular}%
\end{table}

Table \ref{tab:my-table3} presents the detection results of the NF-CSE-CIC-IDS2018-v2 dataset. The ML model has improved the DR of most of the attacks present in the dataset, achieving an accuracy of 96.90\% and an F1 score of 0.98. Most attacks were fully detected with a DR ranging between 99\% to 100\%. However, the detection of certain attack types such as Brute Force, DDoS attack-HOIC, infiltration, and SQL injection is still unreliable when using the extended Netflow feature set. Their respective F1 score is lower due to a high number of false positives. Overall, the performance of the ML model when applied to the NF-CSE-CIC-IDS2018-v2 dataset is superior compared to when using the CSE-CIC-IDS2018 and NF-CSE-CIC-IDS2018 datasets. However, there is an increased prediction time of 27.28 \textmu s compared to 24.17 \textmu s and 17.29 \textmu s, respectively.

\begin{table}[!h]\scriptsize
\centering
\caption{NF-UQ-NIDS-v2 multi-class classification results}
\label{tab:uq}
\begin{tabular}{l|r|r|r|r|}
\cline{2-5}
                                              & \multicolumn{2}{c|}{\textbf{NF-UQ-NIDS}}& \multicolumn{2}{c|}{\textbf{NF-UQ-NIDS-v2}}\\ \hline
\multicolumn{1}{|c|}{\textbf{Class Name}}      & \multicolumn{1}{c|}{\textbf{DR}} & \multicolumn{1}{c|}{\textbf{F1 Score}} & \multicolumn{1}{c|}{\textbf{DR}} & \multicolumn{1}{c|}{\textbf{F1 Score}} \\ \hline
\multicolumn{1}{|l|}{{Analysis}}       & 69.63\%               & 0.21     & 78.43\%               & 0.24         \\ \hline
\multicolumn{1}{|l|}{{Backdoor}}       & 90.95\%               & 0.92     & 89.61\%               & 0.93         \\ \hline
\multicolumn{1}{|l|}{{Benign}}         & 71.70\%               & 0.83     & 93.45\%               & 0.96         \\ \hline
\multicolumn{1}{|l|}{{Bot}}            & 100.00\%              & 1.00     & 100.00\%               & 1.00         \\ \hline
\multicolumn{1}{|l|}{{Brute Force}}    & 99.94\%               & 0.85     & 98.16\%               & 0.74         \\ \hline
\multicolumn{1}{|l|}{{DoS}}            & 55.54\%               & 0.62     & 99.46\%               & 1.00         \\ \hline
\multicolumn{1}{|l|}{{Exploits}}       & 80.65\%               & 0.81      & 85.16\%               & 0.84        \\ \hline
\multicolumn{1}{|l|}{{Fuzzers}}        & 63.24\%               & 0.54     & 80.58\%               & 0.84         \\ \hline
\multicolumn{1}{|l|}{{Generic}}        & 58.90\%               & 0.61     & 85.41\%               & 0.88         \\ \hline
\multicolumn{1}{|l|}{{Infilteration}}  & 60.57\%               & 0.03    & 21.62\%               & 0.19          \\ \hline
\multicolumn{1}{|l|}{{Reconnaissance}} & 88.96\%               & 0.88    & 98.24\%               & 0.76         \\ \hline
\multicolumn{1}{|l|}{{Shellcode}}      & 83.89\%               & 0.15   & 89.35\%               & 0.34           \\
\hline
\multicolumn{1}{|l|}{{Theft}}          & 87.22\%               & 0.15   & 81.66\%               & 0.22           \\\hline
\multicolumn{1}{|l|}{{Worms}}          & 52.97\%               & 0.46   & 87.20\%               & 0.71           \\ \hline
\multicolumn{1}{|l|}{{DDoS}}           & 77.08\%               & 0.69    & 99.43\%               & 1.00          \\ \hline
\multicolumn{1}{|l|}{{Injection}}      & 40.58\%               & 0.50     & 90.03\%               & 0.90         \\ \hline
\multicolumn{1}{|l|}{{MITM}}           & 57.99\%               & 0.10    & 35.97\%               & 0.43          \\ \hline
\multicolumn{1}{|l|}{{Password}}       & 30.79\%               & 0.27      & 97.09\%               & 0.97        \\ \hline
\multicolumn{1}{|l|}{{Ransomware}}     & 90.85\%               & 0.85    & 96.82\%               & 0.87          \\ \hline
\multicolumn{1}{|l|}{{Scanning}}       & 39.67\%               & 0.08     & 97.36\%               & 0.98         \\ \hline
\multicolumn{1}{|l|}{{XSS}}            & 30.80\%               & 0.21    & 95.72\%               & 0.95          \\ \hline
\multicolumn{1}{|l|}{\textbf{Weighted Average}}        & \textbf{70.81\%}               & \textbf{0.79}   & \textbf{96.93\%}               & \textbf{0.97}           \\ \hline
\multicolumn{1}{|l|}{\textbf{Prediction Time (\textmu s)}} & \multicolumn{2}{c|}{\textbf{14.74}}      & \multicolumn{2}{c|}{\textbf{25.67}}                                             \\ \hline
\end{tabular}
\end{table}

Table \ref{tab:uq} compares the attack detection results of the merged NIDS dataset; NF-UQ-NIDS-v2 compared to its former (NF-UQ-NIDS) dataset. Most of the attacks DR has increased by using the extended NetFlow feature set. The detection of attacks such as DoS, Generic, Worms, DDoS, Injection, password, scanning and XSS has significantly improved. However, attacks such as infiltration and MITM have been detected less accurately. Moreover, the time consumed to predict a single test sample has increased from 14.74 \textmu s to 25.67 \textmu s. An increased accuracy from 70.81\% to 96.96\% and an F1 score from 0.79 to 0.97 confirms the enhanced ML model detection capabilities when applied to the extended NetFlow feature set across 20 attack types conducted over several network environments. 

\begin{figure}[!h]
    \centering
    \includegraphics[width=10cm, height=4.5cm]{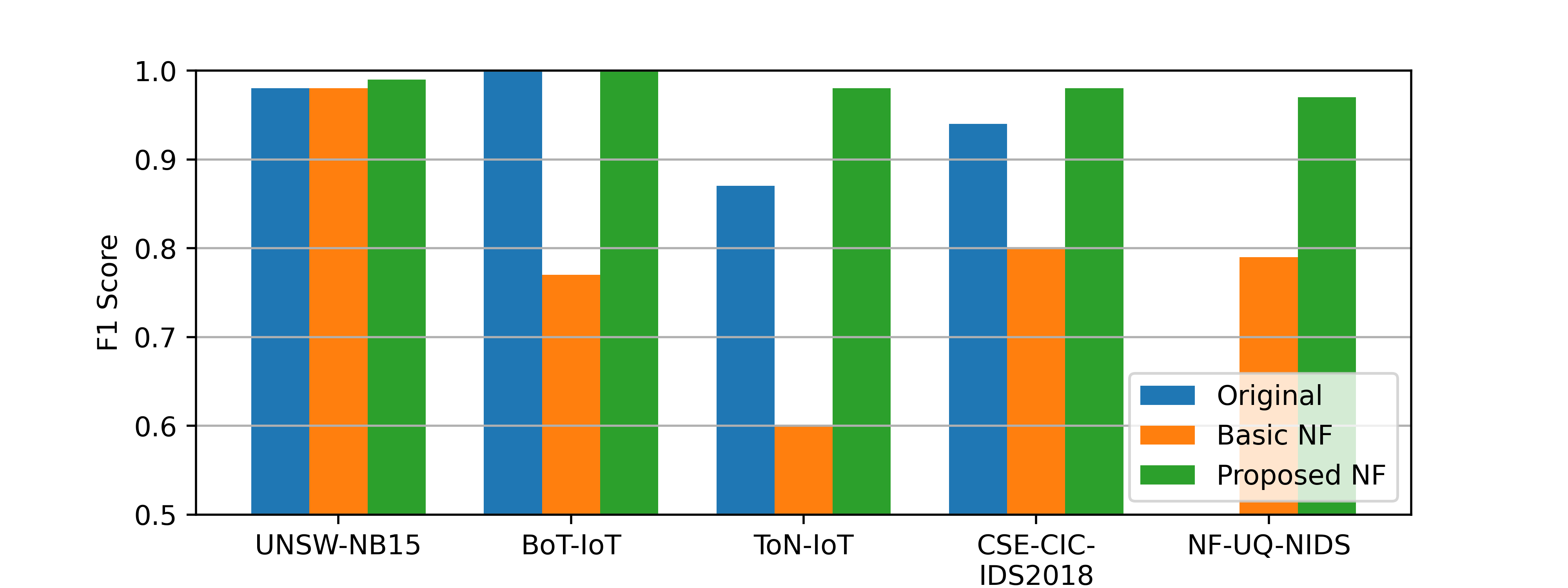}
    \caption{Multi-class classification F1 score}
    \label{we2}
\end{figure}

Overall, the proposed NetFlow feature set has significantly improved the multi-class classification performance of the datasets as displayed in Figure \ref{we2}. The F1 score is plotted on the y-axis and the datasets in their three feature sets are on the x-axis. The detection performance is often comparable to the original feature set but remarkably superior to the basic NetFlow feature set. Hence, the generated datasets enjoy the benefits of adopting a standard common NetFlow feature set and with enhanced detection performance. This motivates the usage of the proposed feature set across future NIDS datasets and encourages researchers to generate their datasets in the proposed format for efficient and reliable ML experiments.

\section{Conclusion}
This paper proposes a NetFlow based standard feature set for NIDS datasets, as listed in  Table \ref{ds}. The importance of having a standard feature set allows the reliable evaluation of ML-based NIDS across multiple datasets, network environments, and attack scenarios. Moreover, the use of a standard feature set allows multiple NIDS datasets to be merged, leading to a larger variety of labelled datasets. As part of the proposed feature set evaluation, five new NIDS datasets have been generated from existing NIDS benchmark datasets. These new dataset variants have been made publicly available to the research community. Our evaluation based on an Extra Tree classifier has shown that our NetFlow-based feature set with 43 features achieves a higher classification performance (F1-Score) than the proprietary feature sets, for all the considered benchmark NIDS datasets, for both binary and multi-class classification scenarios.

The proposed NetFlow-based feature sets have the further advantage of being highly practical and scalable, due to the wide availability of efficient NetFlow exporter and collections. The key benefit of having a standard feature set for NIDS datasets, and the key contribution of this paper, is the ability to more rigorously and reliably evaluate ML-based traffic classifiers across a wide range of datasets, and hence a wider range of attack types, network topologies, etc. This allows the evaluation of how well these ML-based NIDSs can generalise from the dataset they have been trained on, to other network scenarios. We believe the inability to perform such thorough and rigorous evaluation is one of the reasons for the limited deployment of ML-based NIDSs in practical network settings. Therefore, we believe the contributions of this paper can provide a step towards bridging the gap between academic research on ML-based NIDSs and their practical deployment.


\bibliography{main.bib}
\end{document}